\documentclass[smallextended]{svjour3}       
\smartqed  
\usepackage{graphicx}
\usepackage{amsmath}
 \journalname{arXiv}
\begin{document}

\title{On the definition of a concentration function
  relevant to the ROC curve}


\author{Mauro Gasparini   \and
        Lidia Sacchetto 
}


\institute{
  Mauro Gasparini \at
  Department of Mathematical Sciences ``G.L. Lagrange'' \\
Politecnico di Torino \\
Corso Duca degli Abruzzi 24, \\
10124 Torino, Italy \\
              \email{mauro.gasparini@polito.it}           
           \and
Lidia Sacchetto \at
Department of Mathematical Sciences ``G.L. Lagrange'' \\
Politecnico di Torino and Universit\`a di Torino\\
Corso Duca degli Abruzzi 24, \\
10124 Torino, Italy \\
              \email{lidia.sacchetto@polito.it}           
}

\date{Received: date / Accepted: date}

\maketitle

\begin{abstract}
  This is a reader's reaction to a recent paper by
  E. Schechtman and G. Schechtman
  \cite{SS2019} about the correct definition of a concentration
  function for the diagnostic, i.e. supervised classification,
  problem.
  We propose and motivate a different definition and refer to the relevant
  literature.
\end{abstract}

\section{A critical appraisal of a paper by E. Schechtman and G. Schechtman}
\label{intro}
In a paper appeared recently on Metron journal Schechtman and Schechtman
\cite{SS2019} try to shed some light on the relationship between the
Gini Mean Difference (Gini), the Gini Covariance (co-Gini), the Lorenz
curve, the Receiver Operating Characteristic (ROC) curve and a
particular definition of concentration function.  The purpose of the
paper is commendable, since there is a lot of confusion regarding the
various relationships among these concepts.  In particular, we agree
that the ROC curve and its functions (such as the Area Under the
Curve, AUC), as well as an appropriate definition of relative concentration
of a probability distribution with respect to another, are bivariate
objects tying together two different distributions, and can not be
reduced to univariate indices such as the Gini.  Schechtman and
Schechtman \cite{SS2019} build on the wealth of research reviewed in
the monograph by Yitzhaki and Schechtman \cite{YS2013}, where a whole
technology based on the Gini and the co-Gini are proposed as an
alternative to traditional variance and covariance based methods to
study variability, correlation, regression and the like.

Of course, studying  how jointly distributed random variables
interrelate is a very fundamental problem in Statistics
and its applications to Economics and the Sciences.
However, when turning to the diagnostic (or classification) setup,
where ROC curves are naturally used, we observe one or more
diagnostic variables (called features in the Machine Learning
literature) from two populations and try to set up a rule
that discriminates between them.
Some special requirements can then be identified:
\begin{enumerate}
\item Two probability distributions should be evaluated as
  {\em alternative}, mutually exclusive explanations of the data,
  rather than from a {\em joint} point of view; for example,
  a diagnostic marker observed on a patient has either the
  sick patient distribution or the healthy patient distribution,
  and in no way the same marker can be observed jointly under both
  sick and healthy conditions.
\item The definition of the ROC curve and the associated concentration function
  should be viable also in the multivariate setup; for example,
  more than one diagnostic marker can be observed on the same patient.
\item The definition of the ROC curve and the associated concentration function
  must be given both at the population and at the sample level,
  as widely discussed in the ROC literature
  (see for example \cite{KH2009}); a clear definition of the ROC
  curve at the population level is necessary to understand
  basic ideas and to give appropriate definitions.
\end{enumerate}
\renewcommand{\theenumi}{\alph{enumi}}
We claim the definition of (absolute and) relative concentration
curve contained in \cite{SS2019}
is not appropriate for the diagnostic setup since:
\begin{enumerate}
\item conditional distributions are used in the Definition 1 of \cite{SS2019},
  thus contradicting requirement 1);
\item percentiles are used in the same definition, 
  thus contradicting requirement 2);
\item in \cite{SS2019}, the discussion on the ROC curve is mantained
  at the sample level only, making it hard to understand what is, for
  example, the definition of population ROC curve.
\end{enumerate}
We believe the correct definition of concentration function for
the diagnostic setup was given by Cifarelli and Regazzini
in \cite{CR1987} and discussed in \cite{SG2018}, where the
relationship to the ROC curve is also established.  The purpose of
this paper is therefore to show that the definition of concentration
function as given by Cifarelli and Regazzini is more suitable to the
diagnostic setup since it is a one-to-one transformation of the ROC
curve of the optimal diagnostic test, i.e. the one based on the
likelihood ratio.  Some examples are given in Section~\ref{examples}.

\section{The concentration function by Cifarelli and Regazzini and its relationship to the ROC curve of the optimal test}

To favor the comparison with \cite{SS2019}, we adopt a similar notation and
make some simplifying assumptions.
This way, we can state special cases of the main results in
\cite{CR1987} and \cite{SG2018} which possess sufficient generality
to clarify our point but, at the same time, avoid
measure-theoretic complications.

In particular, assume that $Y$ is a continuous random variable
with distribution function $F_Y$ and positive density $f_Y$
and $X$ is a continuous random variable with distribution
function $F_X$ and positive density $f_X$.
$Y$ and $X$ represent, respectively, the relevant diagnostic variable
under the two conditions to be compared by a diagnostic test.
For example, $Y$ may be a biological marker measured in a
diseased person, whereas $X$ is the same marker when measured
in a healthy person.
Next, define the likelihood ratios
$$ L_X =  \frac{f_Y(X)}{f_X(X)} \quad \text{ and } \quad L_Y =  \frac{f_Y(Y)}{f_X(Y)} $$
which are pro bono random variables since they are functions of $X$ and $Y$,
respectively.

\begin{definition}
Assume $L_X$ is continuous with distribution
function $H_X(l)={\rm P}(L_X \leq l)={\rm P}(L_X < l), l>0$
which has an inverse $H^{-1}_X(\cdot)$, its quantile function.
Similarly, assume $L_Y$ is a continuous random variable with distribution
function $H_Y(\cdot)$. Then, the concentration function of $Y$ with respect to
$X$ is defined as the function $\varphi(p), p \in [0,1]$ such that
$\varphi(0)=0$, $\varphi(1)=1$ and
\begin{equation}
\label{defconc}
  \varphi(p) = H_Y(H^{-1}_X(p)), \quad p \in (0,1).
\end{equation}
\end{definition}

The definition is a special case of the one given in \cite{CR1987};
according to their suggestion, for each $p \in [0,1]$ the concentration
function $\varphi(p)$ is the likelihood ratio $Y$-mass of a set collecting the
smallest $p$ fraction of the likelihood ratio $X$-mass.
The strongest simplifying assumption made here is that
$L_X$ and $L_Y$ are continuous random variable; that is not true in
general, even if $X$ and $Y$ are continuous, since partially
parallel densities $f_X$ and $f_Y$ may create atoms in
the distributions of $L_X$ and $L_Y$.

The likelihood ratio is the fundamental measure
of comparison of the $Y$ and $X$ distributions and plays a role similar
(and in our opinion, more appropriate) to the role the conditional expectation
of $Y$ given $X$ plays in \cite{SS2019}.

The likelihood ratio is also a fundamental tool in the definition of the
following optimal diagnostic test in the no-data situation
(i.e. when the $X$ and $Y$ distributions are known and no estimation
is needed)
\begin{definition}
  Suppose an  observation $Z$ has to be assigned
  either to the $X$ or to the $Y$ population.
  Then, for some $0 < t < 1$,  the likelihood ratio based
  diagnostic test assigns $Z$ to $Y$ (resp. $X$) if
  $f_Y(Z)/f_X(Z) > H^{-1}_X(t)$ (resp. $\leq$).
\end{definition}
As it is well known in the ROC literature,
the diagnostic rule based on the likelihood ratio is the best
possible test we can construct based on $Z$, as
discussed for example in \cite{Z2012}. Optimality basically
stems from the Neyman-Pearson lemma.

The point of the above construction is that
the optimal diagnostic test has a ROC curve which is a bijective
transformation of the concentration function in Definition 1 above,
as shown in the following theorem.
\begin{theorem}
\label{theoROC}
  Under the above assumptions, the ROC curve of the optimal likelihood ratio
  based diagnostic test is
  \begin{equation}
    \label{ROC}
  {\rm ROC}_{opt} (q) = 1 - \varphi(1-q) \quad \quad 0 \leq q \leq 1,
  \end{equation}
  while, as usual, ${\rm ROC}_{opt}(0)=0$ and  ${\rm ROC}_{opt}(1)=1.$
\end{theorem}
Proof. As usual, the ROC curve can be calculated as a parametric curve in $t$
by computing separately the True Positive Rate (TPR)
and the False Positive Rate (FPR):
\begin{align*}
TPR(t) &= \text{P} \left( \frac{f_Y(Y)}{f_X(Y)} > H^{-1}_X(t) \right) = 1 - H_Y(H_X^{-1}(t)) \\
FPR(t) &= \text{P} \left( \frac{f_Y(X)}{f_X(X)} > H^{-1}_X(t) \right) = 1 - H_X(H_X^{-1}(t)) = 1-t 
\end{align*}
Setting $q=1-t$ and substituting, we obtain the explicit form (\ref{ROC}).

As a consequence, the ROC of the optimal test is a nondecreasing,
continuous and convex function, while other ROC curves
of suboptimal diagnostic rules may not be.

\section{The Lorenz curve and the AUC of the optimal test}

An interesting special case discussed in \cite{CR1987}
arises when $X$ is a positive random variable
with finite mean $\mu_x= \int_0^{\infty}x f_X(x) dx$
and $Y$ is the {\em length-biased} version of $X$, i.e.
$$
f_Y(y) = \frac{y f_X(y)}{ \mu_X},  \quad y>0.
$$
In economic applications, $Y$ represents wealth;
in general, it may be a transferable character, i.e.
some characteristic which can in theory be transported from one unit
of the population to another. This is the famous Lorenz-Gini setup.
The likelihood ratios in this case simplify to
$$ L_X =  \frac{f_Y(X)}{f_X(X)} = \frac{X f_X(X)}{\mu_X f_X(X)} = \frac{X}{\mu_X}$$ 
and
$$ L_Y =  \frac{f_Y(Y)}{f_X(Y)} = \frac{Y f_X(Y)}{\mu_X f_X(Y)} = \frac{Y}{\mu_X}$$ 
so that $H_X(l) = F_X(\mu_x l)$ and  $H_Y(l) = F_Y(\mu_x l)$
and finally 
$$ \varphi_{Lorenz}(p) = H_Y(H^{-1}_X(p)) = F_Y( F_x^{-1}(p)) =
\frac{\int_0^{F_x^{-1}(p)} y f_X(y) dy}{\int_0^{\infty} x f_X(x) dx},
$$
in which we recognize one of the usual forms of the Lorenz curve.
We have just proven the following
\begin{corollary}
  In the Lorenz-Gini scenario, i.e. when  $ f_Y(y) = y f_X(y) / \mu_X $,
  the concentration curve is the usual Lorenz curve.
\end{corollary}

A second important consequence of Theorem~\ref{theoROC} is about the AUC
of the optimal likelihood ratio based test,
which can be easily computed as follows.
\begin{equation}
\label{AUC}
{\rm AUC}_{opt} = \int_0^1 ROC_{opt}(q) dq
= \int_0^1 (1-\varphi(1-q)) dq = 1-\int_0^1 \varphi(s) ds.
\end{equation}
Now, in the Lorenz-Gini scenario, the Gini concentration coefficient
(Gini) is defined to be twice the area between the diagonal
and the Lorenz curve:
$$
\text{Gini} = 2 \int_0^1 (p - \varphi_{Lorenz}(p)) dp =
1 - 2 \int_0^1 \varphi_{Lorenz}(p) dp
$$
Since the concentration curve is a generalization of the Lorenz curve
which describes the concentration of one variable with respect to another
(and not necessarily its length-biased version),
we can define the generalized Gini as
$$
\text{Gini}_{gen} = 2 \int_0^1 (p - \varphi(p)) dp,
$$
similarly to the co-Gini in \cite{SS2019}. Substituting into
expression~(\ref{AUC}) we obtain the following corollary.
\begin{corollary}
  The AUC of the optimal likelihood ratio based diagnostic test
  equals
  $$
\text{AUC}_{opt} = \frac12{(1+{\rm Gini}_{gen})}.
  $$
\end{corollary}
The same result can be found in \cite{Lee1999} and mentioned by several
other authors. We stress that the result is true for the likelihood ratio
based test and, of course, for models with monotone likelihood ratios
(like the example considered in \cite{Lee1999})
but not in general for the AUC of any ROC, as also noted by \cite{SS2019}.
A few more results we have obtained agree with the results
in \cite{Lee1999}, but they have been presented here in a more general
form at the population level for continuous variables,
for which some examples are presented in the next section.

\section{Some examples}
\label{examples}

{\em Example 1.}
Let $X$ be exponential with rate parameter $\lambda_X$ and
$Y$ be exponential with rate parameter $\lambda_Y$ and assume,
as it is customary, that $\lambda_X > \lambda_Y$, so that $Y$ is
stochastically greater than $X$ (this corresponds to a situation
where the greater a diagnostic marker, the more is indicative
of disease).
Then it is easy to verify that
$$
H_X(l) = \text{P}\left(\frac{f_Y(X)}{f_X(X)} \leq l\right)
= \text{P}\left(\frac{\lambda_Y e^{-\lambda_Y X}}{\lambda_X e^{-\lambda_X X}} \leq l \right)
= 1 - (\frac 1 {rl})^{r/(r-1)}
$$
for $l > 1/r$ and 0 otherwise, where $r=\lambda_X/\lambda_Y$.
Similarly,
$$
H_Y(l) = \text{P}\left(\frac{f_Y(Y)}{f_X(Y)} \leq l\right) = 1 - (\frac 1 {rl})^{1/(r-1)}
$$
for $l > 1/r$ and 0 otherwise. Also,
$$
H_X^{-1}(p) = \frac 1 r (\frac 1 {1-p})^{(r-1)/r}
$$
so that the concentration function is 
$$
\varphi(p) = H_Y(H^{-1}_X(p)) = 1-(1-p)^{1/r}, \quad p \in (0,1),
$$
the ROC curve of the likelihood ratio based optimal test is
$$
{\rm ROC}_{opt} (q) = q^{1/r} \quad \quad 0 \leq q \leq 1,
$$
and
$$
\text{AUC}_{opt} = \frac r {r+1}.
$$

\noindent
{\em Example 2.}
Let $X$ be exponential with rate parameter $\lambda_X$ and assume $Y$
is its length-biased version, so that
$$
f_Y(y) = \frac{y \lambda_X e^{-\lambda_X y}}{1/\lambda_X} = \lambda_X^2 y e^{-\lambda_X y}, \quad y>0,
$$
i.e. $Y$ is a gamma random variable with parameters 2 and $\lambda_X$.
This is a Lorenz-Gini scenario, where it is easy to verify that
$$
H_X(l) = \text{P}(\frac{f_Y(X)}{f_X(X)} \leq l) =\text{P}(\lambda_X X \leq l) =
1 - e^{-l}
$$
whereas, after some calculus,
\begin{multline}
H_Y(l) = \text{P}(\frac{f_Y(Y)}{f_X(Y)} \leq l) = \text{P}(\lambda_X Y \leq l) =
1- e^{-l} -l e^{-l}.
\end{multline}
Since $H_X^{-1}(p) = -\log(1-p)$,
$$
\varphi(p) = p + (1-p) \log(1-p), \quad \text{ROC}_{opt}(q) = q-q\log(q). 
$$

\noindent
{\em Example 3.}
Assume $X$ is normal with mean $\mu_X$ and variance $\sigma^2_X$
and $Y$ is normal with mean $\mu_Y$ and variance $\sigma^2_Y$,
with $\mu_X > \mu_Y$ for the reasons stated in Example 1.
This is the well-know {\em binormal classification model}
which has been studied by several authors.
To compute the concentration and the ROC$_{opt}$ curves in the general case,
one should first compute
$$
H_X(l) = \text{P}(\frac{\sigma_X}{\sigma_Y} {\rm exp}
\{-\frac12(\frac{X-\mu_Y}{\sigma_Y})^2 + \frac12(\frac{X-\mu_X}{\sigma_X})^2\}
\leq l)
$$
and 
$$
H_Y(l) = \text{P}(\frac{\sigma_X}{\sigma_Y} {\rm exp}
\{-\frac12(\frac{Y-\mu_Y}{\sigma_Y})^2 + \frac12(\frac{Y-\mu_X}{\sigma_X})^2\}
\leq l),
$$
a task which can be accomplished by simulation or by tedious
calculations.
Notice that, unlike Examples 1 and 2,
if $\sigma_X \not= \sigma_Y$  the model does not
have monotone likelihood ratios and only the likelihood ratio based
test has a proper ROC curve, as it is well-known in the literature.
If $\sigma_X = \sigma_Y$ instead, the likelihood ratio is a linear function
and therefore monotone.
The two cases generalize to higher dimensions, giving rise
to Fisher's Quadratic and Linear Discriminant functions, respectively.
Further details are contained in \cite{SG2018}.

\section{Conclusions}

The definition of concentration function given here is a convenient
one for the diagnostic problem,
since it compares two alternative probability distributions
using a natural bivariate generalization of the Lorenz curve.
The discussion on the concentration and the ROC curves at the
population level allows for a deeper understanding of the concepts and
for the proof of Theorem 1, which ties together the concentration
function and the ROC curve of the optimal likelihood ratio based
diagnostic test. Similar results were given by \cite{Lee1999}
at the sample level.
All results mentioned in this section can conceptually be generalized to higher
dimensions, although computations may become very hard.
In particular, the likelihood ratio is an efficient
dimension reduction technique which reduces  the comparison
to a one-dimensional problem and allows for     Definition \ref{defconc}
of concentration function without involving higher dimensional
conditional expectations or quantiles.

We hope we have convinced the reader that the nature of the
diagnostic (classification) problem requires a definition of
concentration function which does not involve conditional and joint
distributions of the  populations which are being
compared.

\begin{acknowledgements}
Mauro Gasparini was funded by MIUR, Department of Excellence 2018-2022.
\end{acknowledgements}



\begin{thebibliography}{}

\bibitem{CR1987}
Cifarelli, D.M. and Regazzini, E. 
On a general definition of concentration function. 
{\em SANKHYA B}, 49, 307--319 (1987).

 
\bibitem{KH2009}
Krzanowski, W.J. and Hand, D.J. 
{\em ROC Curves for Continuous Data}, 
Chapman \& Hall, (2009). 

\bibitem{Lee1999}
Lee, W.C.,
Probabilistic analysis of global performances of diagnostic tests:
interpreting the Lorenz curve-based summary measures,
{\em Statistics in Medicine}, 18, 455--471 (1999).

\bibitem{SG2018} 
     Sacchetto, L. and Gasparini, M.,
     Proper likelihood ratio based ROC curves for
  general binary classification problems,
  ArXiv:1809.00694 (2018)

\bibitem{SS2019}
  Schechtman, E. and Schechtman, G.,
  The relationship between Gini terminology and the ROC curve,
  {\em Metron}, 171--178 (2019)
  
\bibitem{YS2013}
  Yitzhaki, S. and  Schechtman, E.,
  {\em The Gini Methodology}. Springer.

\bibitem{Z2012}
Zou K.H., Liu A., Bandos A.I.,
Ohno-Machado L. and Rockette H.E., 
{\em Statistical Evaluation of
Diagnostic Performance
Topics in ROC Analysis
}, Chapman \& Hall (2012)

\end{thebibliography}
\end{document}